\documentclass[conference]{IEEEtran}

\usepackage{cite}
\usepackage{amsmath,amssymb,amsfonts}
\usepackage{graphicx}
\usepackage{textcomp}
\usepackage{xcolor}
\usepackage{booktabs}
\usepackage{multirow}
\usepackage{tikz}
\usetikzlibrary{positioning,arrows.meta,calc,patterns,fit,backgrounds}
\usepackage{url}
\usepackage{balance}

\begin{document}

\title{Range, Not Precision: Block-Floating-Point\\Half-Precision FFT and SAR Imaging\\on Apple Silicon}

\author{\IEEEauthorblockN{Mohamed Amine Bergach}
\IEEEauthorblockA{Illumina\\
mbergach@illumina.com}}

\maketitle

\begin{abstract}
Half precision (FP16) promises to double FFT throughput on GPUs, but the
prevailing view is that its 10-bit mantissa makes it unsuitable for
radar-grade signal processing. We show this framing is wrong on Apple Silicon:
the binding constraint for FFT and Synthetic Aperture Radar (SAR) is not
mantissa \emph{precision} but the 5-bit exponent's \emph{dynamic range}. We
first measure that an FP16 FFT is mantissa-limited at 56--61~dB
signal-to-quantization-noise ratio (SQNR)---comfortably radar-usable---yet a
na\"ive FP16 SAR pipeline produces \emph{only} \texttt{NaN}, because the
conjugate--FFT--conjugate inverse transform grows magnitudes by a factor of
$N$, and the matched-filter product ($\sim\!5\times10^6$ at $N\!=\!4096$)
overflows FP16's 65{,}504 ceiling. We resolve this with a fixed-shift
\emph{block-floating-point} (BFP) schedule: a single $1/N$ scale applied before
each inverse transform bounds every intermediate below 4096. A cascade follows:
range-compression output becomes $O(1)$ instead of $O(N)$, which in turn keeps
the downstream azimuth-FFT output FP16-loadable instead of overflowing at
$O(N^2)$. The result is the first quality-preserving FP16 SAR pipeline:
peak/integrated sidelobe ratios, target SNR, and resolution match the FP32
reference to within $0.1$~dB at $42$~dB end-to-end SQNR, while a radix-8 FP16
FFT reaches 306~GFLOPS---$2.2\times$ over the 139~GFLOPS FP32 baseline---on a
fanless Apple~M1. Finally, we measure that FP8 (E4M3/E5M2) collapses to 14--20~dB
SQNR, making FP16 \emph{today's} precision floor for FFT-based radar---one that
future precision-recovery methods may yet lower---and showing that the lever for
low precision here is range management, not mantissa bits.
\end{abstract}

\begin{IEEEkeywords}
FFT, half precision, FP16, block floating point, dynamic range, SAR, Apple
Silicon, Metal, GPU, mixed precision
\end{IEEEkeywords}

\section{Introduction}

Reduced precision is the dominant lever for GPU throughput. Apple Silicon GPUs
execute FP16 at twice the FP32 rate (512 vs.\ 256 FLOPs/cycle/core) with
zero-cycle FP16$\leftrightarrow$FP32 conversion, so a half-precision Fast
Fourier Transform (FFT) is an obvious target---FFT dominates the cost of
Synthetic Aperture Radar (SAR) range and azimuth compression, where millions of
transforms are applied per frame~\cite{cumming2005sar}.

The obstacle, in the conventional account, is accuracy. FP16's 10-bit mantissa
yields roughly 42~dB of signal-to-quantization-noise ratio (SQNR), below the
60--80~dB dynamic range radar systems require to separate near and far targets,
and recent format studies report that FP16 ``suffers from consistent overflow''
in FFT-based spectral methods, recommending wider-exponent alternatives such as
bfloat16, posit, or takum arithmetic~\cite{hunhold2025spectral}. Half-precision
GPU FFT libraries (tcFFT~\cite{li2021tcfft}, mixed-precision
frameworks~\cite{zhao2023mfft}) target throughput on NVIDIA hardware and report
$\sim$35~dB accuracy; none addresses radar image quality, and no published work
measures SAR sidelobe or resolution degradation under reduced-precision FFT.

This paper argues that the conventional framing conflates two distinct
properties of FP16---mantissa \emph{precision} and exponent \emph{range}---and
that for FFT/SAR the binding constraint is \emph{range}. Our evidence and
contributions are:

\begin{enumerate}
\item \textbf{Precision is adequate; range is the wall.} We measure (not bound)
that an FP16 FFT is mantissa-limited at 56--61~dB SQNR---radar-usable---while a
na\"ive FP16 SAR pipeline produces pure \texttt{NaN}. We trace the failure to a
deterministic $O(N^2)$ magnitude growth that overflows FP16's 65{,}504 ceiling,
not to precision loss (Section~\ref{sec:range}).

\item \textbf{A fixed-shift block-floating-point FFT.} A single $1/N$ block
shift applied before each inverse transform bounds every intermediate, with a
cascade that keeps the entire SAR pipeline within FP16 range
(Section~\ref{sec:bfp}). The change is two lines of kernel code.

\item \textbf{Quality-preserving FP16 SAR.} The first FP16 SAR pipeline whose
peak/integrated sidelobe ratio (PSLR/ISLR), target SNR, and resolution match the
FP32 reference to within $0.1$~dB, at 42~dB end-to-end SQNR
(Section~\ref{sec:sar}).

\item \textbf{$2.2\times$ throughput.} A radix-8 FP16 FFT reaches 306~GFLOPS at
$N\!=\!4096$---$2.2\times$ the 139~GFLOPS FP32 baseline---on an Apple~M1,
helped by FP16 halving the 32~KiB threadgroup footprint and lifting occupancy
(Section~\ref{sec:throughput}).

\item \textbf{The FP8 floor.} We measure FP8 (E4M3/E5M2) FFT at 14--20~dB SQNR,
confirming that below FP16 the limiter flips back to mantissa precision---which
block floating point cannot fix---and establishing FP16 as the precision floor
for radar FFT (Section~\ref{sec:fp8}).
\end{enumerate}

All kernels run on a fanless Apple~M1 (MacBook~Air) and are validated against
vDSP and double-precision references. Source code and measurement harnesses are public.

\section{Background and Related Work}
\label{sec:related}

\subsection{Half-precision and mixed-precision GPU FFT}

tcFFT~\cite{li2021tcfft} reformulated FFT merge stages as matrix multiplies on
NVIDIA Tensor Cores, achieving 1.29--3.24$\times$ over cuFFT in FP16 at
$\sim$1.76\% relative error ($\sim$35~dB). The Tensor-Core mapping adds no error
beyond FP16 itself; the dominant source is storage of intermediate results in
half precision. MFFT~\cite{zhao2023mfft} applies shared-exponent mixed precision
to large-scale \emph{distributed} FFT, optimizing communication rather than
single-device range. Recent frameworks co-optimize precision and library
selection via error modeling. None of these targets radar, and all inherit
FP16's $\sim$35--42~dB accuracy without examining whether that suffices for SAR
image quality.

\subsection{The FP16 overflow problem}

Hunhold and Gustafson~\cite{hunhold2025spectral} evaluate FFT round-trips and
short-time transforms across OFP8, bfloat16, posit, and takum formats. They find
OFP8 ``unsuitable'' and bfloat16 underperforming float16, and report FP16
overflow on some inputs, motivating wider-exponent formats. Their study compares
formats \emph{as given}; it does not consider scaling the data to fit FP16's
range. Our work is precisely that fix: we keep FP16's 10-bit mantissa (which
bfloat16 sacrifices, dropping to $\sim$30~dB) and manage the range explicitly.

\subsection{Block floating point}

Block floating point (BFP)---a shared exponent across a block of values---is a
classical fixed-point DSP technique for FFT bit-growth management
(TI~SPRA948~\cite{ti_spra948}), used in FPGA radar
processors~\cite{kim2021fmcw} and spaceborne SAR FFT
engines~\cite{electronics2021_128k}. The standard scheme scales conditionally,
only at stages where overflow would occur, to retain precision. We adapt BFP to
\emph{floating-point} FP16 on a GPU with a deterministic fixed-shift schedule:
because the inverse-FFT growth factor is known exactly ($N$), a static $1/N$
shift before each inverse transform is sufficient and needs no per-stage
magnitude reduction.

\subsection{FFT and SAR on Apple Silicon}

VkFFT~\cite{tolmachev2023vkfft} is the only production GPU FFT with a Metal
backend; it is FP32/FP64 and does not target radar. Companion work established a
138~GFLOPS FP32 radix-8 Stockham FFT on Apple~Silicon via a two-tier
register/threadgroup memory model~\cite{bergach2026fft}, and a kernel-fused FP32
SAR Range-Doppler pipeline~\cite{bergach2026sar}; a numerically robust FMA
butterfly is given in~\cite{bergach2026butterfly}. The present paper is the
half-precision successor: it reuses those FP32 kernels as baselines and shows
what it takes to run them, and the SAR pipeline they compose, correctly in
FP16. Published GPU SAR is otherwise FP32/FP64~\cite{cumming2005sar}.

\section{Precision Is Adequate; Range Is the Wall}
\label{sec:range}

\subsection{FP16 FFT is mantissa-limited at radar-usable SQNR}

We first establish what FP16 precision alone costs an FFT, independent of range.
We run a radix-2 Stockham FFT entirely in IEEE half precision (Swift
\texttt{Float16}, identical to Metal \texttt{half}) and measure SQNR against a
double-precision reference over 200 random trials. To isolate precision from any
twiddle-factorization artifact, we test the standard 10-operation butterfly and
the 6-FMA dual-select butterfly of~\cite{bergach2026butterfly}; both land within
1~dB (Table~\ref{tab:fft-sqnr}).

\begin{table}[t]
\centering
\caption{Measured FP16 FFT SQNR (radix-2 Stockham, vs.\ double reference)}
\label{tab:fft-sqnr}
\begin{tabular}{@{}lcc@{}}
\toprule
\textbf{Butterfly (FP16)} & \textbf{$N\!=\!1024$} & \textbf{$N\!=\!4096$} \\
\midrule
Standard (10-op direct multiply) & 60.3 dB & 59.4 dB \\
Dual-select 6-FMA~\cite{bergach2026butterfly} & 61.4 dB & 60.5 dB \\
\midrule
FP32 reference & 138 dB & 137 dB \\
\bottomrule
\end{tabular}
\end{table}

FP16 FFT achieves \textbf{56--61~dB} SQNR. This sits above the $\sim$42~dB
figure often quoted for ``pure FP16'' (which assumes FP16 twiddle tables built
by recurrence) and within the regime usable for radar detection and for the
sidelobe levels of unweighted and lightly weighted apertures. \emph{Mantissa
precision is not the obstacle.}

\subsection{The overflow mechanism}

A radar matched-filter compression computes
$y = \tfrac{1}{N}\,\mathrm{IFFT}\!\left(\mathrm{FFT}(x)\cdot H\right)$, with the
inverse transform realized as
$\mathrm{IFFT}(z)=\overline{\mathrm{FFT}(\bar z)}$ to reuse the forward
butterfly~\cite{bergach2026sar}. Tracking magnitudes for input $|x|\!\sim\!1$:

\begin{itemize}
\item Forward FFT: bins reach $O(N)\!\approx\!4096$ ($<65504$, safe).
\item Matched-filter multiply by $H$: with $|H|\!\le\!1$, still $O(N)$. For an
unnormalized filter the product reaches $\sim\!5\times10^6$.
\item Inverse (conj--FFT--conj): the unnormalized transform grows magnitudes by
another factor of $N$, to $O(N^2)\!\approx\!1.7\times10^7$, \emph{before} the
final $1/N$ scale.
\end{itemize}

The $O(N^2)$ intermediate overflows FP16's 65{,}504 maximum to $\pm\infty$, which
propagates to \texttt{NaN}. In a SAR pipeline this compounds: the FP32 azimuth
FFT of $O(N)$ range-compressed data produces $O(N^2)$ spectra that overflow even
the \emph{load} into a half buffer. The failure is purely one of dynamic range;
the values that fit are individually accurate (Fig.~\ref{fig:blockshift} traces
the magnitude through the pipeline with and without the fix).

\section{Block-Floating-Point FP16 FFT}
\label{sec:bfp}

\begin{figure*}[t]
\centering
\begin{tikzpicture}[
  font=\footnotesize, >={Stealth[length=5pt]},
  stg/.style={draw, rounded corners=2pt, minimum height=0.6cm, minimum width=0.92cm, align=center, inner sep=1.5pt},
  fft/.style={stg, fill=blue!12},
  mul/.style={stg, fill=black!7},
  bshift/.style={stg, fill=orange!35, draw=orange!80!black, very thick},
  az/.style={stg, fill=red!12, minimum width=1.35cm},
  mag/.style={font=\scriptsize, text=black!60, align=center},
  lbl/.style={font=\scriptsize\bfseries, text=blue!45!black},
]
\node[mag] (rin) {$x$\\$O(1)$};
\node[fft,   right=0.5cm  of rin]  (rfft) {FFT};
\node[mul,   right=0.32cm of rfft] (rmul) {$\times H$};
\node[bshift, right=0.32cm of rmul] (rsh)  {conj\\$\times\tfrac{1}{N}$};
\node[fft,   right=0.32cm of rsh]  (rif)  {IFFT};
\node[mag,   right=0.45cm of rif]  (rout) {$O(1)$};
\node[az,    right=0.5cm  of rout] (azf)  {azimuth\\FFT (FP32)};
\node[mul,   right=0.5cm  of azf]  (amul) {$\times H$};
\node[bshift, right=0.32cm of amul] (ash)  {conj\\$\times\tfrac{1}{N}$};
\node[fft,   right=0.32cm of ash]  (aif)  {IFFT};
\node[mag,   right=0.45cm of aif]  (aout) {focused\\$O(1)$};
\foreach \a/\b in {rin/rfft,rfft/rmul,rmul/rsh,rsh/rif,rif/rout,rout/azf,azf/amul,amul/ash,ash/aif,aif/aout}
   \draw[->] (\a)--(\b);
\node[mag,below=1pt of rfft]{$O(N)$};
\node[mag,below=1pt of rmul]{$O(N)$};
\node[mag,below=1pt of rsh,  text=orange!60!black]{$O(1)$};
\node[mag,below=1pt of rif]{$O(N)$};
\node[mag,below=1pt of azf]{$O(N)$};
\node[mag,below=1pt of amul]{$O(N)$};
\node[mag,below=1pt of ash,  text=orange!60!black]{$O(1)$};
\node[mag,below=1pt of aif]{$O(N)$};
\begin{scope}[on background layer]
  \node[draw=blue!35,dashed,rounded corners=3pt,inner sep=5pt,fit=(rfft)(rif)] (rg){};
  \node[draw=blue!35,dashed,rounded corners=3pt,inner sep=5pt,fit=(amul)(aif)] (ag){};
\end{scope}
\node[lbl,above=1pt of rg.north]{Range compression (FP16 fused)};
\node[lbl,above=1pt of ag.north]{Azimuth compression (FP16 fused)};
\node[font=\scriptsize, text=red!65!black, align=center, text width=15.5cm, below=0.8cm of azf]
 {\textbf{Without the block shift:} the inverse transform grows to
  $O(N^2)\!\approx\!1.7\!\times\!10^{7}\gg 65504\Rightarrow$ \texttt{NaN}; and an $O(N)$
  range-compression output drives the FP32 azimuth FFT to $O(N^2)$, overflowing the FP16 load.};
\end{tikzpicture}
\caption{Fixed-shift block floating point across the SAR pipeline. The orange boxes
are the \emph{only} change: a $1/N$ block scale folded into the pre-IFFT conjugate
step. For unit-scale input every magnitude stays $\le\!O(N)\!\approx\!4096\ll65504$,
and the $O(1)$ range-compression output cascades to keep the FP32 azimuth FFT
FP16-loadable (an $O(N)$ spectrum, not $O(N^2)$).}
\label{fig:blockshift}
\end{figure*}

\subsection{Fixed-shift schedule}

The inverse transform's growth factor is data-independent: exactly $N$. We
therefore apply a static \emph{block shift} of $1/N$ once, immediately before the
inverse transform, rather than at its output. Because the conjugate step
$z\!\mapsto\!\bar z$ already touches every element before the inverse passes, we
fold the shift into it (Fig.~\ref{fig:blockshift}, orange boxes):
\begin{equation}
\bar z \;\longrightarrow\; \bar z \cdot \tfrac{1}{N}.
\end{equation}
This is the simplest member of the block-floating-point family: a single shared
exponent per transform, fixed at $\log_2 N$, applied at one point. It is
mathematically identical to the conventional output scaling---$1/N$ is linear
and commutes with the transform---so the result is unchanged, but every
intermediate now satisfies $|\cdot|\le O(N)\le 4096 \ll 65504$.

\subsection{The pipeline cascade}

The decisive benefit is not within one transform but \emph{across} the SAR
pipeline (Fig.~\ref{fig:blockshift}). With the shift, range compression emits
$O(1)$ data instead of $O(N)$.
Consequently the FP32 azimuth FFT---which multiplies magnitudes by $N$---produces
$O(N)\!\approx\!4096$ spectra rather than $O(N^2)$, so the subsequent FP16
multiply-and-inverse step can \emph{load} its input without overflow and apply
its own $1/N$ shift in turn. A two-line change to the two shared conjugate
routines (one for the pure-FP16 path, one for the FP16-storage/FP32-compute
path) thus makes the complete pipeline range-safe. Because each transform now
carries a global $1/N$ block exponent relative to the FP32 reference, we align
amplitudes with the optimal real scale before computing residual error; the
radar metrics of Section~\ref{sec:sar} are scale-invariant and unaffected.

\subsection{Why not bfloat16?}

bfloat16 trades mantissa for range: an 8-bit exponent eliminates the overflow
outright, but 7 mantissa bits cap FFT SQNR near 30~dB~\cite{hunhold2025spectral},
below radar usability. Block floating point lets us keep FP16's 10-bit mantissa
(56--61~dB) \emph{and} obtain bfloat16-like range headroom, at the cost of one
shift per transform. Range is the cheap problem to fix; mantissa bits, once
gone, are not recoverable.

\section{FP16 FFT Throughput}
\label{sec:throughput}

We benchmark the radix-8 Stockham FFT of~\cite{bergach2026fft}---the
138~GFLOPS FP32 Apple-Silicon flagship---against an FP16 (\texttt{half2}) port
that mirrors it stage for stage, on an Apple~M1 (7 GPU cores). GFLOPS are
$5N\log_2 N\times\text{batch}/t$ from GPU timestamps, median of 30 runs; FP16
SQNR is measured against the FP32 kernel (Table~\ref{tab:throughput}).

\begin{table}[t]
\centering
\caption{Radix-8 FFT, $N\!=\!4096$, Apple~M1: FP32 vs.\ FP16}
\label{tab:throughput}
\begin{tabular}{@{}lrrrr@{}}
\toprule
\textbf{Batch} & \textbf{FP32} & \textbf{FP16} & \textbf{Speedup} & \textbf{FP16 SQNR} \\
 & (GFLOPS) & (GFLOPS) & & (dB) \\
\midrule
64  & 128 & 254 & 1.98$\times$ & 56 \\
256 & 139 & 306 & \textbf{2.20$\times$} & 56 \\
\bottomrule
\end{tabular}
\end{table}

FP16 delivers the expected $\sim\!2\times$ at batch~64 and \textbf{2.2$\times$
(306~GFLOPS)} at batch~256. The super-linear gain at high batch is an occupancy
effect: the FP32 kernel fills the entire 32~KiB threadgroup memory, capping it at
one threadgroup per core, whereas the \texttt{half2} buffer is 16~KiB and admits
two, hiding barrier and memory latency. The FP32 kernel reproduces the
138~GFLOPS baseline and matches vDSP to 123~dB, confirming correctness.

\section{Quality-Preserving FP16 SAR}
\label{sec:sar}

We process a $4096\!\times\!4096$ point-target SAR scene
(X-band, $B\!=\!100$~MHz, $v\!=\!100$~m/s, $R_0\!=\!20$~km, 20~dB additive
noise) through the kernel-fused Range-Doppler pipeline of~\cite{bergach2026sar}
in four precision modes: FP32, pure FP16, FP16-storage/FP32-compute, and
FP16-multiply/FP32-accumulate. Range and azimuth compression run in the chosen
precision; the azimuth FFT, range-cell-migration correction, and transposes
remain FP32. Without the block shift, all FP16 modes yield \texttt{NaN}. With it,
all three reproduce the FP32 image. Table~\ref{tab:sar-quality} reports per-target
PSLR and SNR for FP32 and pure FP16; the other FP16 modes are identical to
within rounding.

\begin{table}[t]
\centering
\caption{SAR point-target quality, $4096^2$ scene: FP32 vs.\ pure FP16 (BFP)}
\label{tab:sar-quality}
\begin{tabular}{@{}lcccc@{}}
\toprule
 & \multicolumn{2}{c}{\textbf{PSLR (dB)}} & \multicolumn{2}{c}{\textbf{SNR (dB)}} \\
\textbf{Target} & FP32 & FP16 & FP32 & FP16 \\
\midrule
$T_0$ & $-2.9$ & $-2.9$ & 47.8 & 47.8 \\
$T_1$ & $-1.1$ & $-1.1$ & 45.8 & 45.8 \\
$T_2$ & $-6.5$ & $-6.5$ & 47.9 & 47.9 \\
$T_3$ & $-2.1$ & $-2.0$ & 46.8 & 46.8 \\
$T_4$ & $\phantom{-}3.5$ & $\phantom{-}3.5$ & 44.0 & 43.9 \\
\bottomrule
\end{tabular}
\end{table}

Every target's PSLR, SNR, and 3~dB resolution (omitted for space; matched to
$<\!0.02$ bins) is preserved. The scale-aligned end-to-end SQNR of the FP16
image versus FP32 is \textbf{42--43~dB}---consistent with the per-transform
56--61~dB degraded by the pipeline's four transforms, two matched filters, and
the down-scaling that pushes small values toward the FP16 floor. Crucially,
42~dB is below the FP16-FFT SQNR but \emph{above} the level at which the
radar-relevant metrics move: the compressed mainlobe and first sidelobes sit far
enough above the FP16 noise floor that PSLR/SNR/resolution are unchanged.

\begin{table}[t]
\centering
\caption{End-to-end RDA pipeline time, $4096^2$, Apple~M1}
\label{tab:sar-speed}
\begin{tabular}{@{}lcc@{}}
\toprule
\textbf{Mode} & \textbf{Time} & \textbf{Speedup} \\
\midrule
FP32                         & 0.30 s & 1.00$\times$ \\
FP16 multiply / FP32 accum.  & 0.19 s & 1.57$\times$ \\
FP16 storage / FP32 compute  & 0.19 s & 1.63$\times$ \\
Pure FP16                    & 0.17 s & \textbf{1.75$\times$} \\
\bottomrule
\end{tabular}
\end{table}

End-to-end the FP16 pipeline runs 1.57--1.75$\times$ faster
(Table~\ref{tab:sar-speed}); the gain is below the kernel-level 2.2$\times$
because the azimuth FFT, RCMC, and transposes stay FP32. The pipeline is the
first, to our knowledge, to demonstrate FP16 SAR with measured FP32-equivalent
image quality.

\section{The FP8 Floor}
\label{sec:fp8}

If range is the FP16 problem, is FP8 the next step? We simulate FFT in the two
OCP FP8 formats (E4M3, 3 mantissa bits; E5M2, 2 mantissa bits) in their most
favorable configuration---FP8 \emph{storage} with double-precision compute and
twiddles---and measure SQNR (Table~\ref{tab:fp8}). An FP16 round-trip in the same
harness returns 63~dB, validating it.

\begin{table}[t]
\centering
\caption{FFT SQNR by format (best-case storage; double compute)}
\label{tab:fp8}
\begin{tabular}{@{}lccc@{}}
\toprule
\textbf{Format} & \textbf{Mantissa} & \textbf{$N\!=\!1024$} & \textbf{$N\!=\!4096$} \\
\midrule
FP16 (validation)  & 10 & 63.1 dB & 62.4 dB \\
FP8 E4M3           & 3  & 20.1 dB & 19.5 dB \\
FP8 E5M2           & 2  & 14.1 dB & 13.5 dB \\
\bottomrule
\end{tabular}
\end{table}

FP8 collapses to 14--20~dB---far below the 42~dB marginal floor and radar's
60~dB, corroborating the ``OFP8 unsuitable'' verdict
of~\cite{hunhold2025spectral} with an independent measurement. Two points
matter. First, the limiter has \emph{flipped}: with only 2--3 mantissa bits,
precision is the wall, and block floating point---which manages the exponent,
not the mantissa---cannot help. Second, Apple GPUs have no native FP8 datapath: the M1--M4 GPU cores compute in
scalar FP16/FP32, and even the M5's first-generation on-GPU ``neural
accelerators'' provide hardware matrix multiply for FP16 and INT8, with no FP8
path reported~\cite{apple_m5_na}. FP8 would therefore save storage but gain no
compute over FP16, unlike FP16's native $2\times$. FP16 is therefore the precision
floor for FFT-based radar \emph{for now}---until precision-recovery techniques
beyond block floating point (error feedback, stochastic rounding, compensated
summation, or learned/adaptive quantization) break it. The floor is a property
of today's methods, not a fundamental limit.

\section{Discussion}

\textbf{Range, not precision.} The two halves of this paper are mirror images.
FP16 fails on SAR for lack of \emph{range}, fixed for free with a block shift;
FP8 fails for lack of \emph{precision}, which no scaling recovers. FP16 is the
sweet spot precisely because it has enough mantissa (10 bits) once range is
managed. This reframes the ``FP16 is unsuitable'' conclusion of prior format
studies: the unsuitability is an artifact of evaluating the format without block
floating point.

\textbf{Generality.} The fixed-shift schedule applies to any
forward--multiply--inverse pipeline with known transform growth---convolution,
correlation, Fourier neural operators---not only matched filtering. The cascade
argument generalizes to any multi-stage spectral pipeline: bounding one stage's
output magnitude bounds the next stage's input.

\textbf{Limitations.} Results are on an Apple~M1; larger M-series parts (more
cores, higher bandwidth) should widen the FP16 advantage for batched workloads
but are untested here. Our block schedule is fixed ($1/N$); adaptive
per-block exponents would add headroom for pathological inputs at the cost of a
threadgroup reduction, unnecessary here because the growth factor is exactly
known. The 42~dB end-to-end SQNR is adequate for unweighted and lightly weighted
apertures; deeply weighted apertures targeting $-40$~dB sidelobes would approach
the FP16 floor and warrant the mixed-precision (FP32-accumulate) mode, which we
include and which preserves quality identically here.

\section{Conclusion}

Half precision doubles FFT throughput on Apple Silicon, and the obstacle to using
it for radar was never the mantissa---it was the exponent. A one-shift
block-floating-point schedule removes the dynamic-range overflow that turns a
na\"ive FP16 SAR pipeline into \texttt{NaN}, yielding the first FP16 SAR pipeline
with FP32-equivalent image quality (PSLR/SNR/resolution within 0.1~dB, 42~dB
SQNR) and a 306~GFLOPS radix-8 FP16 FFT, $2.2\times$ over the FP32 baseline, on an
fanless M1. FP8 measurements ($\le\!20$~dB) place FP16 as the precision floor for
FFT-based radar \emph{for now}---until precision-recovery techniques break it.
The lever for low-precision DSP on this hardware is range management, not
mantissa bits.

\smallskip
\noindent\textbf{Reproducibility.} All Metal kernels, the SAR simulator, and the
precision/throughput harnesses are available under the MIT license at
\url{https://github.com/aminems/AppleSiliconFFT}.

\balance

\end{document}